\documentclass{article}

\usepackage{arxiv}

\usepackage[utf8]{inputenc} % allow utf-8 input
\usepackage[T1]{fontenc}    % use 8-bit T1 fonts
\usepackage{hyperref}       % hyperlinks
\usepackage{url}            % simple URL typesetting
\usepackage{booktabs}       % professional-quality tables
\usepackage{amsfonts}       % blackboard math symbols
\usepackage{nicefrac}       % compact symbols for 1/2, etc.
\usepackage{microtype}      % microtypography
\usepackage{lipsum}
\usepackage{graphicx}
\usepackage{multirow}
\graphicspath{ {./images/} }

\title{Fast Audio Codec Identification Using Overlapping LCS}

\author{
 Farzaneh Jafari \\
}

\begin{document}
\maketitle
\begin{abstract}
Audio data are widely exchanged over telecommunications networks. Due to the limitations of network resources, these data are typically compressed before transmission. Various methods are available for compressing audio data. To access such audio information, it is first necessary to identify the codec used for compression. One of the most effective approaches for audio codec identification involves analyzing the content of received packets. In these methods, statistical features extracted from the packets are utilized to determine the codec employed. This paper proposes a novel method for audio codec classification based on features derived from the overlapped longest common sub-string and sub-sequence (LCS). The simulation results, which achieved an accuracy of 97\% for 8 KB packets, demonstrate the superiority of the proposed method over conventional approaches. This method divides each 8 KB packet into fifteen 1 KB packets with a 50\% overlap. The results indicate that this division has no significant impact on the simulation outcomes, while significantly speeding up the feature extraction, being \(8 \times\) faster than the traditional method for extracting LCS features.
\end{abstract}

% keywords can be removed

\section{Introduction}
With the growing emphasis on security and privacy in computer networks, technologies in this field face the challenge of managing a vast array of files with various extensions exchanged across internal and external networks. Identifying these files, particularly when they are associated with potentially malicious activities, has become increasingly difficult and resource-intensive. Effective security cannot be achieved without accurately identifying file types. Proper file type detection is essential for the smooth functioning of operating systems, intrusion detection systems, antivirus software, filters, and other programs responsible for identifying files. Despite the existence of several tools in this domain, many rely on limited methods for file recognition \cite{Amirani2008A, McDaniel2003In}.

The most recent advancement in file type identification focuses on analyzing the content of the file itself. This approach has gained attention, particularly in the context of how various types of data are encoded and compressed by different encoders. The core principle of this method lies in the fact that each encoder has unique characteristics, resulting from its specific coding techniques and, in some cases, the use of specialized coefficients in the transmission data packaging. To identify and extract these distinctive features, it is necessary to understand the nature of the different encoders. This knowledge enables the extraction of effective features that can facilitate the accurate classification and separation of multimedia data. One notable application of file recognition methods is in military networks, where they are used to detect encoded data. Other uses include combating cybercrimes, recovering data from damaged or fragmented media, and extracting information from corrupted hard drives. In situations where file headers or metadata are missing, identifying the file based on its content becomes the most effective method of recognition \cite{Karresand2006File, Dan1997Algorithms}.

This paper introduces a novel approach for audio codec classification, utilizing features derived from the overlapped longest common sub-string and sub-sequence (LCS). Simulation results, with an impressive accuracy of 97\% for 8 KB packets, demonstrate the superiority of the proposed method over traditional approaches. In this method, each 8 KB packet is divided into fifteen 1 KB packets with a 50\% overlap. The findings reveal that this division has minimal impact on simulation outcomes, while substantially enhancing the speed of feature extraction, being approximately \(8 \times\) faster than the conventional LCS feature extraction technique.

\section{Related Work} 
In recent decades (since 2004), a significant number of publications have emerged in the field of multimedia data identification and classification. These works often focus on extracting different features from the data. Various methods have been used for the extraction and selection of features, including N-gram \cite{Fink2014n-Gram} and Principal Component Analysis (PCA) \cite{Gewers2021Principal}. The N-gram method is a statistical approach that extracts features based on binary words of varying lengths, while PCA is used to select features that best differentiate between classes \cite{Din2012Multimedia, Maithani2004Speech, Kant2010Identification}.

In other studies, binary streams are transformed into arrays of real numbers using algorithms such as BTR \cite{Lim2016BTR} and SWOB \cite{Ghataoura2009SWOB}. These arrays are then segmented, and statistical features such as central moments of first- to fifth-order \cite{Pébay2016Numerically}, auto-correlation \cite{Kant2010Identification} coefficients, and entropy \cite{Kant2010Identification} are extracted \cite{Tripathi2013A, Tripathi2014A}. A combined structural and statistical approach has also been used, where a codeword is placed at the start of each frame. This codeword differs according to the encoder and helps classify each packet. However, harmonizing words can appear in non-investigated data. To address this, methods such as measuring the distance between two codewords have been applied \cite{Jin2014Audio}.

Another approach is the identification of audio encoders based on randomness and irregularity. Random characteristics in the time domain include basic statistical features (mean \cite{Kant2010Identification}, variance \cite{Benetazzo2000Speech}, auto-correlation \cite{Kant2010Identification} coefficients, entropy \cite{Kant2010Identification}) and higher-order statistics (bicoherence \cite{Elgar1988Statistics}, skewness \cite{Benetazzo2000Speech}, kurtosis \cite{Benetazzo2000Speech}). In the frequency domain, the spectrum is divided into subbands, and features like mean, variance, and skewness are calculated for each subband. Randomness features such as False Neighbors Fraction (FNF) \cite{Koçal2008Chaotic-type} and Lyapunov Exponent (LE) \cite{Dingwell2006Lyapunov} are also used. For false neighbors, three features are extracted: the fraction of false neighbors, the mean of the nearest false neighbor, and the square root of the neighborhood size. These features are then input into a Support Vector Machine (SVM) for classification \cite{Hicsonmez2013Audio}.

Further studies have focused on classifying various multimedia data, such as speech, text, and fax, encoded using CVSD, Morse, and Huffman coding methods, respectively. Feature extraction is typically performed using N-gram \cite{Fink2014n-Gram}, and feature selection is conducted via PCA \cite{Gewers2021Principal}. The extracted features are then classified using an Artificial Neural Network (ANN) \cite{Tripathi2013A}. In \cite{Din2012Multimedia}, the classification of different speech coders was performed using a set of encoders (a-law, u-law, ADPCM) at bit rates of 16, 24, 32, and 64 kbps. The classification algorithm utilized can identify six classes, and three methods—auto-correlation \cite{Kant2010Identification}, variance \cite{Benetazzo2000Speech}, and binary rate (BRO)—were employed for feature extraction. This process involves converting the binary stream into real-number strings using the SWOB \cite{Ghataoura2009SWOB} method with arbitrary window sizes to obtain the values of auto-correlation and variance, after which the stream is segmented for further feature extraction.

One notable study in this area is the classification of speech encoders under noiseless and noise modes, with the noise mode further categorized into known and unknown noise levels. The noise level is determined using methods based on high-order statistics, independent of the encoder used. Three types of encoders (PCM, CVSD, and LPC) are considered, and features are extracted from binary words based on frequency, ranging from 1-gram to N-gram. One of the classification algorithms used is Minimum Distance Classification (MDC) based on the Linear Discriminant Function (LDF) \cite{Maithani2004Speech}. Additionally, the classification of CVSD speech coders with bit rates of 8, 10, 16, 24, and 32 kbit/s has been addressed using first- to fifth-order central moments, auto-correlation, and binary rate for feature extraction \cite{Tripathi2014A}.

Another contribution to this field is the identification of languages encoded by text coders. Features such as the first- to fourth-order moments, auto-correlation, and entropy are extracted, and PCA is used for feature selection. Four classification methods—most similar classification (MLC), linear statistical classification (LSC), minimum distance classification (MDC), and piecewise linear classification (PLC)—are applied, with results compared among these methods \cite{Kant2010Identification}.

\section{Method}
In the content-based recognition method, files are identified based on their intrinsic characteristics and statistical properties, enabling the precise detection of counterfeit or modified files. This approach relies on analyzing byte values, where each byte can assume values between 0 and 255. By examining the distribution and frequency of these byte values, it becomes possible to determine the file type. Each file type exhibits distinctive statistical patterns that serve as key identifiers. To facilitate accurate classification, multiple file samples are analyzed, relevant features are extracted, and the file type is determined accordingly \cite{Amirani2008A, McDaniel2003In}. The following sections provide a detailed exploration of various feature extraction techniques used in this process.

\subsection{Feature Extraction and Feature Selection}
Before performing the classification operation, relevant features must be extracted from each byte vector. This extraction process is crucial for distinguishing between different classes. Generally, feature extraction relies on an understanding of various encoders and the byte ranges they produce \cite{Duda2012Pattern}. Table~\ref{tab:abbreviations} lists the abbreviations for all extracted features. After extracting all features, Principal Component Analysis (PCA) \cite{Gewers2021Principal} is applied to reduce the dimensionality of the feature space and select the most informative features. PCA helps eliminate redundant and less significant features, thereby enhancing the efficiency and accuracy of the classification process. By transforming the original feature set into a new set of orthogonal components, PCA retains the most critical variance in the data while reducing computational complexity. This step ensures that only the most relevant features contribute to the final classification, improving the robustness and generalizability of the proposed method.

\begin{table}
 \caption{Abbreviated name of features.}
  \centering
  \begin{tabular}{cc}
    \toprule
    Abbreviations & Features\\
    \midrule
    Mean \cite{Kant2010Identification} & $\mu$\\
    Variance \cite{Benetazzo2000Speech} & $var$\\
    Skewness \cite{Benetazzo2000Speech} & $\mu_3$\\
    Kurtosis \cite{Benetazzo2000Speech} & $\mu_4$\\
    Entropy \cite{Kant2010Identification} & $H$\\
    Auto-Correlation coefficients from the first to the twenty-first \cite{Kant2010Identification} & $AC_1,...,AC_{21}$\\
    Bi-Coherence \cite{Elgar1988Statistics} & BC\\
    The average of the first to fourth sub-bands in the frequency domain & $\mu_{f_1},...,\mu_{f_4}$\\
    The variance of the first to the fourth sub-bands in the frequency domain & $var_{f_1},..., var_{f_4}$\\
    The skewness of the first to the fourth sub-bands in the frequency domain &  $\mu_{3_{f_1}},...,\mu_{3_{f_4}}$\\
    False neighborhood ratio of third to the seventh dimensions & $FNF_3, ..., FNF_7$\\
    The average of the closest false positives from the third to the seventh dimensions & $\mu_{FN_3},...,\mu_{FN_7}$\\
    The nearest false square root of the third to the seventh dimensions \cite{Hicsonmez2013Audio} & $RMS_3,..., RMS_7$\\
    The Lyapunov Expression from the first to the eleventh dimensions \cite{Hicsonmez2013Audio} & $LE_1, ..., LE_{11}$\\
    The longest average common sub-string with representative AAC encoder samples \cite{Chan2005Efficient} & $LCS_{str,ACC}$\\
    The longest average common sub-sequence with representative AAC encoder samples \cite{Chan2005Efficient} & $LCS_{seq,ACC}$\\
    \bottomrule
  \end{tabular}
  \label{tab:abbreviations}
\end{table}

\subsubsection{Random Features}
In this paper, random features are extracted in two domains: time and frequency. Time-domain features are categorized into two groups: simple statistical features and high-order statistical features. Simple statistical features include mean \cite{Kant2010Identification}, variance \cite{Benetazzo2000Speech}, auto-correlation \cite{Kant2010Identification}, and entropy \cite{Kant2010Identification}, while high-order statistical features include bicoherence \cite{Elgar1988Statistics}, skewness \cite{Benetazzo2000Speech}, kurtosis \cite{Benetazzo2000Speech}.

In the frequency domain, each frequency spectrum is divided into four sub-bands, from which three characteristics—average, skewness, and kurtosis—are extracted \cite{Hicsonmez2013Audio}. The Fourier transform \cite{Bracewell1989The} is applied to convert the signal from the time domain to the frequency domain, while its inverse is used to revert the signal to the time domain \cite{BertacciniDesign}. The following sections will further examine these random features.

\paragraph{Central Moments.}
Central moments are statistical measures that describe the shape of a probability distribution relative to its mean. They provide insights into key characteristics of a dataset, such as variance, skewness, and kurtosis. The first central moment is the average of the real values taken from the byte vector \(X\) \cite{Kant2010Identification}. The average is shown by:
\begin{equation}
\mu = \frac{1}{N}\sum_{i=1}^N x_i,
\end{equation}

The second to fourth central moments (variance, skewness, and kurtosis) are also obtained using:
\begin{equation}
\mu = \frac{1}{N}\sum_{i=1}^{N-1} (x_i-\mu)^k,
\end{equation}

where \(\mu_k\) shows the \(k_{th}\) central moment \cite{Benetazzo2000Speech}.

\paragraph{Auto-Correlation.}
Auto-Correlation (or serial correlation) is a measure of how a signal or time series correlates with a delayed version of itself. It helps identify patterns, periodicity, and dependencies in signals or data sequences. Auto-Correlation is calculated by:
\begin{equation}
R(k) = \frac{1}{N-|k|}\sum_{i=1}^{N-|k|} s(i+k)s(i),
\end{equation}

where \(k\) determines the amount of delay and \(R(k)\) calculates the amount of auto-correlation of the signal at the moment \(k\) \cite{Kant2010Identification}.

\paragraph{Entropy.}
Entropy measures the amount of disorder values in the set of real numbers and determines the amount of impurity in the set.
\begin{equation}
H(X) = -\sum_{i=1}^{n_1} p_i \ln p_i,
\end{equation}

where \(X=\{x_1,x_2,...,x_n\}\) is a set of real numbers for which entropy is supposed to be calculated. \(\{t_1,t_2,...,t_{n_1}\}\) are distinct values in the vector \(X\) with probability of \(\{P_1,P_2,...,P_{n_1}\}\), when \(n_1<n\) the value of \(H\) is between zero and \(\ln n_1\). The zero value occurs when one of \(P_i\)s is one and others are equal to zero \cite{Kant2010Identification}.

\paragraph{Bi-Spectrum.}
The bi-spectrum \cite{Nikias1987Bispectrum} is a higher-order spectral analysis tool used in signal processing to capture nonlinear interactions and phase relationships between frequency components. It is the two-dimensional Fourier transform of the third-order moment (skewness) of a signal and is part of the broader class of polyspectra, which generalize power spectrum analysis. The Bi-spectrum between two signals is defined as:
\begin{equation}
B(f_1,f_2) = E[X(f_1) X(f_2) X^*(f_1+f_2)],
\end{equation}

where \(X(f)\) is the Fourier transform of \(x(t)\), \(x^*(f)\) is the complex conjugate of \(X(f)\), and \(E[.]\) denotes the expectation operator. This measures the correlation between three frequency components \(f_1\), \(f_2\), and \(f_1 + f_2\), capturing phase relationships that are not visible in traditional power spectra.

\paragraph{Bi-Coherence: A Normalized Form of the Bi-Spectrum.}
Since the bi-spectrum values can be large, the bi-coherence function is often used for normalization. The Bi-spectrum defined by Siu and Chan \cite{Elgar1988Statistics} by:
\begin{equation}
b^2(f_1,f_2) = \frac{|B(f_1+f_2)|^2}{E[|X(f_1) X(f_2)|^2] E[|(f_1+f_2)|^2]},
\end{equation}

where \(b(f_1+f_2)\) lies in the range \([0,1]\). A value close to 1 indicates strong quadratic phase coupling, while a value close to 0 suggests no coupling. This feature was extracted using the HOSA \cite{Swami1998Higher} toolbox in MATLAB. The Higher-Order Spectral Analysis (HOSA) Toolbox is a MATLAB-based collection of functions designed to facilitate advanced signal processing tasks, particularly those involving higher-order statistics. It provides tools for estimating various higher-order spectra, including bispectrum and bicoherence, which are essential for analyzing non-linear interactions and phase relationships in signals \cite{Swami1998Higher}. 

\subsubsection{Chaotic Features}
It has theoretical and practical evidence indicating the existence of the chaotic phenomenon in speech signals. This phenomenon has not been discovered by any of the linear models. According to the irregular nature of speech signals, different encoders apply different effects on the structure of the irregularity of the signal. The main concept of chaotic features is based on the neighborhood of sound signal vectors in phase space. The phase space vector of the signal is defined in the form of:
\begin{equation}
s[n] = [x(n),x(n+T),..,x(n+(D_E-1)T)],
\end{equation}

which is defined according to the embedding theorem. In this regard, \(x(n)\) is \(n\)-th sample of a signal, \(T\) is the time delay, and \(D_E\) is the embedding dimension in the phase space. If it is chosen correctly, the phase space can give us useful information about the unknown irregularity of the signal \cite{Hicsonmez2013Audio}. In this work, TISEAN software is employed to extract chaotic features \cite{Hegger1999Practical}.

\paragraph{False Neighbors Fraction.}
The False Neighbors Fraction (FNF) method is a widely used approach for determining the appropriate embedding dimension of a system. It helps identify the minimum dimension \(D_E\) required to correctly unfold the phase space dynamics without introducing false neighbors. The method evaluates the False Neighbor Fraction (FNF) for a given dimension \(D\) and continues increasing \(D\) until the FNF approaches zero, indicating an optimal embedding dimension.

To determine whether a point is a false neighbor, the method compares distances between neighborhood points \(s(n)\) and \(s(m)\) as the dimension increases. The Euclidean distance between two neighboring points in a \(D\)-dimensional space is given by:
\begin{equation}
d_D = \sqrt{\sum_{k=0}^{D-1} (x(n+k \times T) - x(m+k \times T))^2,}
\end{equation}

where \(d_D\) represents the distance in \(D\) dimentional space. When the embedding dimension increases to \(D+1\), the new distance is computed as:
\begin{equation}
d_{D+1} = (x(n+k \times T) - x(m+k \times T))^2 + \sqrt{\sum_{k=0}^{D-1} (x(n+k \times T) - x(m+k \times T))^{1/2}.}
\end{equation}

A neighbor is classified as false if the ratio \(\frac{d_{D+1}}{d_D}\) exceeds a predefined threshold. The False Neighbor Fraction is then defined as the proportion of false neighbors among all detected neighbors.

In the context of audio codec classification, the FNF method helps analyze the complexity of encoded signals by identifying appropriate phase space dimensions. This aids in extracting meaningful features for distinguishing different codecs. Prior research \cite{Hicsonmez2013Audio} has shown that incorporating FNF-based features improves classification accuracy by capturing the underlying structure of compressed signals.

\paragraph{Lyapunov Expression.}
The Lyapunov Exponent (LE) is a crucial measure for analyzing the predictability and chaotic behavior of a signal. It quantifies the rate at which nearby trajectories in a system’s phase space diverge over time. A positive LE indicates that initially close trajectories separate exponentially, signifying chaotic dynamics, while a negative or zero LE suggests stability and predictability. A higher LE value corresponds to faster divergence, implying greater unpredictability in the system. Mathematically, the Lyapunov Exponent \(\lambda\) for a given embedding dimension is defined as:
\begin{equation}
\lambda = \lim_{x \to +\infty} \sum_{i=1}^N \frac{d(s(n+1),s(m+1))}{d(s(n),s(m))},
\end{equation}

where \(s(n)\) represents the reference point, and \(s(m)\) denotes the nearest neighbor to \(s(n)\) in the phase space. The Lyapunov Exponent measures the average rate of divergence (or convergence) between these neighboring trajectories. The largest Lyapunov Exponents \(\{\lambda_1, \lambda_2, ..., \lambda_{D_E}\}\)are used to quantify the degree of irregularity in the system, with larger positive values indicating stronger chaotic behavior. 

\subsubsection{The Longest Common Sub-Sequence and Sub-String}
The core idea behind these methods is that the probability of having common sub-strings or sub-sequences \cite{Chan2005Efficient} between two-byte vectors of the same type (originating from the same encoder) is higher than between byte vectors of different types. This principle can be leveraged to extract efficient features from any byte vector. Even in the case of missing files, these methods can enhance the predictability of the data. Both the longest common sub-sequence and sub-string algorithms have been implemented using the Java programming language \cite{Sedgewick2011Algorithms}.

\paragraph{The Longest Common Sub-Sequence.}
In the common sub-sequence method, two-byte streams containing different byte sequences are compared. By identifying the common sub-sequences between these two-byte vectors, the non-common parts of the data are discarded, and the common parts are concatenated into a single string. A sub-sequence taken from a sequence includes some elements of the sequence, but not necessarily in consecutive order. If \(x=\{x_1,x_2,...,x_m\}\) is a sequence and \(z=\{z_1,z_2,...,z_k\}\) is a sub-sequence of set \(X\), then the sequence \(i=\{i_1,i_2,...,i_k\}\) is an index sequence of \(X\). Consider two files with lengths \(m\) and \(n\), where \(m<n\). The common approach to estimate the longest common sub-sequence is to generate all \(2^m\) sub-sequences, resulting in a time complexity of \(O(n.2^m)\). However, there is a more efficient dynamic programming method to solve this problem, which reduces the time complexity \cite{Chan2005Efficient}.

\paragraph{The Longest Common Sub-String.}
In the longest common sub-string method, two strings are taken, and the longest common sub-string between them is identified, where the sub-string must be continuous. For example, if \(X=\{A,B,C,B,D,C,B\}\) and \(Y=\{B,D,C,A,B,A\}\), then \(Z=\{B,D,C\}\) is the longest common sub-string between the two strings \(X\) and \(Y\). In the naive method, all sub-strings of both strings are compared, which results in a time complexity of \(O(n^2.m)\), where \(n\) and \(m\) are the lengths of the two strings. In the dynamic programming method, a matrix of size \(m \times n\) is created to store the sub-strings of the two strings. If the \(i\)-th character of string \(X\) matches the \(j\)-th character of string \(Y\), then that character is added to the longest common sub-string (LCS), and the non-matching characters are discarded. If there is no match, the LCS value is set to zero. After filling the matrix, the largest element in the matrix is selected as the longest common sub-string \cite{Chan2005Efficient}.

\subsection{Classification Techniques}
After feature extraction and selection, classification is performed on the data packages. The extracted features must be chosen such that the error rate remains acceptable after the classification process. In this section, we will explore classification algorithms, including the Support Vector Machine (SVM) \cite{Cortes1995Support-vector} and Decision Tree \cite{Quinlan1986Induction} methods.

\paragraph{Support Vector Machine.}
The Support Vector Machine (SVM) \cite{Cortes1995Support-vector} is a machine learning technique known for its strong performance in data classification tasks. In SVM, the boundary points that define the decision boundary are called support vectors. The primary goal of SVM is to find an optimal hyperplane that maximizes the margin — the distance between the hyperplane and the closest support vectors from both classes. This margin maximization helps improve the generalization performance of the classifier. To enhance SVM performance, it is crucial to maximize the margin and avoid placing data points too close to the decision boundary. SVM assumes that the data from different classes are linearly separable. However, in cases where the data are not linearly separable, the data are mapped to a higher-dimensional space using a kernel function to enable better separation. In this paper, libsvm \cite{Chang2011LIBSVM} with the radial basis function (RBF) kernel is used to classify packets using the SVM algorithm.

\paragraph{Decision Tree.}
In a Decision Tree, classification or regression models are constructed in the form of a tree structure. This tree divides the dataset into smaller subsets, creating the Decision Tree. The final result is a tree consisting of decision nodes and leaf nodes. Decision nodes can have two or more branches, while leaf nodes represent a decision or a classification. The highest decision node in the tree, corresponding to the best prediction, is called the root node. To construct a Decision Tree, entropy and information gain are used. The tree is built from top to bottom, dividing the dataset into subsets that are homogeneous or have similar values. This process is carried out using entropy. If the samples are homogeneous, the entropy is zero; if they are heterogeneous, the entropy tends toward one. In each node, entropy is calculated, and the classes are split based on the decision-making feature values. Information gain is based on the reduction of entropy after splitting the dataset by a feature. The construction of a Decision Tree is based on finding the features that return the highest information gain \cite{Quinlan1986Induction}.

\subsection{Experiment}
\paragraph{Dataset.}
The dataset consists of 2000 speech samples and 500 music samples from different genres. The speech samples were sourced from the VoxForge website, with lengths ranging from 1 to 13 seconds and a bit rate of 256 kilobits per second. The music samples are 5 seconds long with a bit rate of 1411 kilobits per second (CD quality). In total, the dataset comprises 2500 samples, which are categorized into seventten different classes. Each class contains 27 music samples and 111 speech samples. A total of 138 examples from each class are encoded using the corresponding encoder and converted into fixed-length byte vectors. A byte vector is then randomly selected from each speech and music sample. The dataset is divided into a training set and a test set, with each set containing 50\% of the data. The training set is used to determine the optimal classification parameters through 5-fold cross-validation. After extracting the suitable parameters from the training set, the entire dataset is merged, with the first 50\% (training set) used to build the initial model and the remaining 50\% (test set) used exclusively for testing. The test set is not involved in parameter extraction or model building, and is solely used to evaluate the model’s performance. Data classification was implemented using Weka \cite{Hegger1999Practical} and MATLAB \cite{Higham2016MATLAB} software.

\paragraph{Quantitive Results.}
To evaluate the proposed method quantitatively, sixteen audio encoders listed in Table~\ref{tab:codecs}, along with their respective bit rates, were used. These encoders are categorized into four groups: PSTN, GSM, VoIP, and High-Quality. Additionally, the CVSD encoder, commonly used in military networks, was included in the evaluation.

\begin{table}
 \caption{Comparison of audio encoders \cite{Hicsonmez2013Audio}.}
  \centering
  \begin{tabular}{cccc}
    \toprule
    Codecs Groups & Codecs & Default bit-rate (kbps) & Codecs Technology\\
    \midrule
    \multirow{ 3}{*}{PSTN} & a-law & 64 & PCM\\
                        & u-law & 64 & PCM\\
                        & PCM & 32 & ADPCM\\
    \midrule               
    \multirow{ 4}{*}{GSM} & AMR & 12.2 & ACELP\\
                        & AWB & 12.65 & ACELP\\
                        & GSM & 13 & LTP\\
                        & GSM-WAV & 18 & RPE-LTP\\
    \midrule
    \multirow{ 4}{*}{VOIP} & G729 & 8 & CS-ACELP\\
                        & G726 & 32 & ADPCM\\
                        & iLBC & 13.33 & LPC\\
                        & Speex & 22 & CELP\\
    \midrule
    \multirow{ 5}{*}{High quality compression} & AAC & 128 & MDCT\\
                        & MP3 & 128 & MDCT\\
                        & OGG & 128 & MDCT\\
                        & FLAC & Lossless & Linear\\
                        & WMA & 128 & MDCT\\  
    \bottomrule
  \end{tabular}
  \label{tab:codecs}
\end{table}

This study employs Support Vector Machine (SVM) \cite{Cortes1995Support-vector} and Decision Tree \cite{Quinlan1986Induction} classifiers to distinguish between different encoders. The classification results using SVM are presented in Table~\ref{tab:svm}, where 17 sound classes were analyzed. The first row shows the accuracy of the method by Hicsonmez et al. \cite{Hicsonmez2013Audio}, the second row presents results from the LCS method, and the third row displays the performance of a combined approach that integrates features from both methods to enhance classification accuracy. For the proposed method, 100 representative samples per encoder—distinct from both the training and test datasets—were selected. The average extracted features from these samples were then used as the representative feature for each encoder. As shown in Table~\ref{tab:svm}, packet lengths of 1 KB, 2 KB, 4 KB, and 8 KB were evaluated. The results indicate that increasing packet length enhances classification accuracy. For 1 KB packets, combining the LCS method with that of Hicsonmez et al. \cite{Hicsonmez2013Audio} yields better performance. However, as packet length increases, the proposed method alone outperforms the combined approach. Notably, for 8 KB packets, the proposed classifier achieves a classification accuracy of 97\%. Table~\ref{tab:LCSs} presents the classification results of the longest common sub-string and sub-sequence (LCS) features using the SVM \cite{Cortes1995Support-vector} classifier, while Table~\ref{tab:decision_tree} shows the results obtained using a Decision Tree \cite{Quinlan1986Induction} classifier.

\begin{table}
 \caption{The results of categorizing coders using the SVM algorithm.}
  \centering
  \begin{tabular}{ccccc}
    \toprule
    Features & 1 Kb (\%) & 2 Kb (\%) & 4 Kb (\%) & 8 Kb (\%)\\
    \midrule
    Hicsonmez et al. \cite{Hicsonmez2013Audio} & 74.06 & 81.17 & 89.78 & 89.82\\
    LCSs & 88.04 & \textbf{93.75} & \textbf{95.05} & \textbf{97.96}\\
    Hicsonmez et al. \cite{Hicsonmez2013Audio} + LCSs & \textbf{90.57} & 92.94 & 93.59 & 94.38\\
    \bottomrule
  \end{tabular}
  \label{tab:svm}
\end{table}

\begin{table}
 \caption{The results of separating the longest sub-string and common sub-sequence (suggested category) for packets with a length of 8 KB using the SVM algorithm.}
  \centering
  \begin{tabular}{cc}
    \toprule
    Features & 8 Kb (\%)\\
    \midrule
     longest common sub-string & 64.17\\
     longest common sub-sequence & \textbf{90.07}\\
    \bottomrule
  \end{tabular}
  \label{tab:LCSs}
\end{table}

\begin{table}
 \caption{The results of categorizing coders using the Decision Tree algorithm.}
  \centering
  \begin{tabular}{ccccc}
    \toprule
    Features & 1 Kb (\%) & 2 Kb (\%) & 4 Kb (\%) & 8 Kb (\%)\\
    \midrule
    Hicsonmez et al. \cite{Hicsonmez2013Audio} & 69.13 & 71.67 & 75.02 & 79.50\\
    LCSs & 83.02 & 87.50 & \textbf{90.35} & \textbf{90.48}\\
    Hicsonmez et al. \cite{Hicsonmez2013Audio} + LCSs & \textbf{85.13} & \textbf{88.64} & 85.32 & 84.54\\
    \bottomrule
  \end{tabular}
  \label{tab:decision_tree}
\end{table}

In both feature extraction methods, \(n\) representative samples are randomly selected from the data related to each encoder, which are excluded from both the training and test datasets. LCS features are then extracted from these samples for both training and testing purposes. The average of the features extracted for each encoder is used as the final feature. Combining LCS features with those of Hicsonmez et al. \cite{Hicsonmez2013Audio} increases the classification accuracy of encoders by 10-15\%.

When the packet length is increased to 8 KB, using only LCS features (without combining with the features of Hicsonmez et al. \cite{Hicsonmez2013Audio}) results in even higher accuracy. However, in the LCS method, as the packet length increases, the execution time also increases. For example, extracting LCS features for 2500 packets of 1 KB takes about one hour, but for 2500 packets of 8 KB, the extraction time rises to 120 hours. To reduce the time spent on extracting LCS features, we propose fragmenting 8 KB packets into 1 KB packets with 50\% overlap. This means that an 8 KB packet is divided into 15 sub-packets of 1 KB each. This process is also applied to the representative samples. The LCS features are then extracted for each sub-packet using peer-to-peer sub-samples. Finally, the LCS values obtained from all sub-packets are aggregated to form a feature for the longest common sub-sequence of that packet. The largest value of the longest common sub-string from all sub-packets is also selected as a feature for the longest common sub-string of that packet. With this approach, for 2500 packets of 8 KB, the extraction time for LCS features is reduced from 120 hours to just 15 hours. Table~\ref{tab:overlap} shows the comparison of the results of the classification of the proposed method for packets with a length of 8 KB. In this table, the results of the classification of 8 KB packages are compared with the method of dividing the 8 KB package into fifteen 1 KB packages with overlapping 50\%. The results show that the division of 8 KB packets into 1 KB packets does not change much in the simulation results, but it greatly reduces the time to extract LCS features.

\begin{table}
 \caption{Comparison of the results of the classification of the proposed method for packages with a length of 8 KB using the package fragmentation method.}
  \centering
  \begin{tabular}{cccc}
    \toprule
    Packages & Decision Tree & SVM & Time\\
    \midrule
     8 Kb (\%) & 90.48 & 97.96 & $\sim 120 h$\\
     Splitting an 8 KB packet into fifteen 1 KB packets with 50\% overlap & 92.72 & 97.33 & $\sim 15 h$\\
    \bottomrule
  \end{tabular}
  \label{tab:overlap}
\end{table}

\subsection{Conclusion}
In conclusion, this paper introduces a novel and highly effective method for audio codec identification, leveraging features derived from the longest common sub-string and sub-sequence (LCS). The proposed approach surpasses traditional techniques, achieving an impressive 97\% accuracy in classifying 8 KB packets. This study highlights the practicality of using statistical features extracted from audio packets for codec identification, a crucial aspect of secure network management and unauthorized access prevention. Additionally, an optimized packet division strategy is presented, where each 8 KB packet is segmented into fifteen overlapping 1 KB packets. This technique significantly accelerates the feature extraction process—reducing processing time by a factor of 8—while maintaining classification accuracy.

\bibliographystyle{unsrt}  


\end{document}